\documentclass[aps,prb,twocolumn,superscriptaddress,showpacs,floatfix]{revtex4}

\usepackage{graphicx}
\usepackage{epsfig}
\usepackage{psfrag}
\usepackage{amsmath}

\begin{document}

\title{Adiabatic optical entanglement between electron spins in separate quantum dots}

\author{S. K. Saikin}
\email[]{saykin@fas.harvard.edu} \affiliation{Department of
Physics, University of California, San Diego, La Jolla, CA 92093}
\affiliation{Department of Physics, Kazan State University, Kazan
420008, Russian Federation}
\author{C. Emary} \affiliation{Institut
f\"ur Theoretische Physik, TU Berlin, Hardenbergstr. 36, D-10623,
Germany}
\author{D. G. Steel}
\affiliation{Department of Physics, The University of Michigan Ann
Arbor, MI 48109-1040}
\author{L. J. Sham}
\affiliation{Department of Physics, University of California, San
Diego, La Jolla, CA 92093}

\date{\today}

\begin{abstract}
We present an adiabatic approach to the design of entangling
quantum operations with two electron spins localized in separate
InAs/GaAs quantum dots via the Coulomb interaction between
optically-excited localized states. Slowly-varying optical pulses
minimize the pulse noise and the relaxation of the excited states.
An analytic ``dressed state'' solution gives a clear physical
picture of the entangling process, and a numerical solution is
used to investigate the error dynamics. For two vertically-stacked
quantum dots we show that, for a broad range of dot parameters, a
two-spin state with concurrence $C>0.85$ can be obtained by four
optical pulses with durations $\sim 0.1 - 1$~ns.
\end{abstract}



\maketitle

Adiabatic passage uses the slow variation of a system's
Hamiltonian to select a particular quantum path while avoiding
unintended dynamics. Controlled adiabatic evolution of the ground
state has been proposed as a model for quantum
computation.\cite{Farhi} Stimulated Raman adiabatic passage
(STIRAP) \cite{STIRAP} can be used to transfer populations or
coherences between quantum states through a ``dark state'' which
efficiently suppresses relaxation. Arbitrary single-qubit
operations can be produced, for example, by STIRAP in a tripod
system \cite{kis} or adiabatically controlled Raman excitation in
a $\Lambda$-system.\cite{Clive1} In this work we study how
adiabatic control can be used in design
of optically-induced two-qubit quantum operations.

In systems with a permanent interaction between qubits, it is
known that adiabatic passage through degenerate dressed states can
also be used to construct two-qubit entangling
gates.\cite{Unanyan} However, for scalable solid-state quantum
computation, it is important to keep the qubits isolated from each
other except during gating. Electron spins in semiconductor
quantum dots (QDs) are promising candidates for just such
qubits.\cite{QDQC} They have long coherence time,\cite{Greilich}
can be manipulated by electric gates \cite{gatedQDs} or
optically,\cite{imam,pier} and the coupling between the qubits can
be induced externally.

Significant experimental and theoretical effort has been invested
in optical manipulation of electrons in single and coupled
semiconductor QDs. Schottky diode structures with embedded
self-assembled QDs have been designed to control the number of
electrons in the dots by adjusting the external bias voltage.
\cite{Warburton} The particular optical transitions between the
charged and the excitonic states can be addressed in these dots by
frequency and polarization selection. \cite{Xiaodong}  Efficient
spin-initialization schemes have been demonstrated recently using
optical pumping in the Faraday \cite{Atature} (magnetic field
parallel to the optical axis) and the Voigt \cite{Xiaodong}
(magnetic field orthogonal to the optical axis) configurations.
The Faraday \cite{Readout} and the Kerr \cite{Awschalom} rotations
from single spins confined in QDs have been observed, which should
allow spin-readout and single-spin rotation operations. For
two-qubit quantum operations the energy level structure and the
interdot coupling in vertically-aligned QD pairs have been
studied.\cite{Gerardot,Gammon}

Several designs of two-qubit gates have been recently proposed
utilizing, for example, tunneling between excited states of
QDs,\cite{Clive2QD}, F\"{o}rster-type interaction \cite{Briggs05}
long-range coupling through a photon bus,\cite{Yamamoto} and
electrostatic coupling between the excited states.\cite{Zoller0,
Rossi} These schemes are yet to demonstrated experimentally,
however. The major difficulties are:

\noindent - The proposals utilize properties of the QDs or device
structures which do not exist yet. For instance, two-qubit gates
in Ref.~\onlinecite{Yamamoto} utilize QDs in cavities coupled to a
common waveguide. Though, such a design could potentially allow
large spatial separation of the qubits there are no reliable
device structures yet.

\noindent - The interdot coupling via, for example, electron
tunneling between the excited orbitals, or a F\"{o}rster-type
interaction requires precise alignment of the energy levels and
cannot be controlled experimentally at the present stage of
technology.

\noindent - Demonstration of a two-qubit operation is complicated
because of the gate structure. Though, mathematically all the
two-qubit entangling gates are equivalent, their physical
realization, demonstration and implementation into a particular
quantum algorithm require different amount of resources. It is
particularly important when the operational noise is a main
limiting factor. For instance, demonstration of conditional phase
operation additionally involves a number of single qubit gates
that themselves are very noisy and require a substantial
experimental effort.

In this study we present a general approach to the design of
two-qubit entangling operations with {\it uncoupled} electron
spins in semiconductor QDs utilizing the Coulomb interaction of
transient optically-excited states localized in the dots. We show
that adiabatic pulses combined with the counter-intuitive pulse
ordering of STIRAP allows the construction of non-local two-spin
unitary transformations, whilst efficiently suppressing population
transfer out of the qubit subspace.
Compared to other two-qubit gates with spins in semiconductor QDs our proposal:

\noindent - utilizes the conventional Schottky barrier device
structures within which QDs are routinely grown;

\noindent - is based on the Coulomb interaction between the
excited electronic states in different dots, and therefore does
not require precise control for the energy level structure;

\noindent - provides flexibility in the gate design. In addition
to the control phase gates one can construct operations resulting
in a coherent oscillation of two-spin state population, which is 
a more accessible signature of entanglement.

As illustration we describe an operation for two spins in
separate self-assembled InAs/GaAs QDs. While for clarity, the
entangling process is described in the path language, it, in fact,
represents a quantum operation, made up of a product of
$\sqrt{i{\rm SWAP}}$ and controlled-phase gates. Combined with
single qubit rotations \cite{Clive1} and optical initialization,
\cite{Clive, Xiaodong} we obtain a set of gates for universal
quantum computation. We employ the Voigt configuration to obtain
the flexibility required to select the desired quantum paths
through polarization and frequency selection.  The evolution of
the system is then guided through a particular subset of quantum
paths by a sequence of adiabatic pulses. In our dressed-state
picture the scheme can be viewed as an adiabatic passage of an
arbitrary initial two-spin state through two long-lived states.
The interference between the two paths results in an effective
rotation in the spin subspace. The method proposed here can be
adapted to construct CPHASE and CNOT gates.

\begin{figure}
\centering
\includegraphics[width=0.85\linewidth]{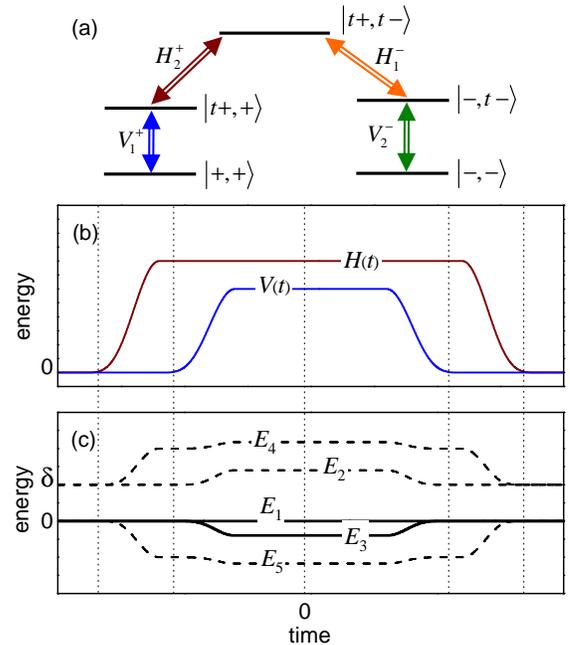}
\caption{(Color online) (a) Optical scheme to control the
entanglement between spins in two InAs/GaAs QDs in the Voigt
configuration. Two-dot states are denoted by kets such as
$|t+,+\rangle$, with  $|\pm \rangle$ for the spin states and $|t\pm
\rangle$ the trion states. Arrows indicate the linear
polarizations $V_j^\pm$ and $H_j^\pm$ for the transitions
$|\pm\rangle \leftrightarrow |t\pm\rangle$ and $|\pm\rangle
\leftrightarrow |t\mp\rangle$ of dot $j= 1,2$. (b) Timing
of pulses for either dot. $V(t)$
and $H(t)$ are envelope functions, for which we use the same shape, rectangular with
fronts shaped as $\sin^4(\pi t/T_f)$, for all pulses, and
the same amplitudes for both V-pulses and  for both H-pulses. (c)
Adiabatic time evolution of the dressed state energies. Solid lines 
show the essential energies which drive the operation.} \label{fig_1}
\end{figure}

In two self-assembled InAs/GaAs QDs, the direct electron or hole
tunneling between the dots may be suppressed by selecting the dot
heights and the interdot distance. \cite{Gerardot,Clive2QD} Then,
because the electrons and holes are confined differently, the
intrinsic Coulomb coupling between particles in different dots
modifies the optical transition energies. \cite{Gerardot,Gammon}
We employ this phenomenon to perform two-qubit operations.
This is similar to the dipole blockade. \cite{Jaksch} However, we
do not rely on an external electric field. This substantially
simplifies the experimental setup and makes the operation less
sensitive to external noise than the proposal of
Ref.~\onlinecite{Zoller0} in which in-plane gates were used.
The particular path used for the entangling operation is shown in
Fig.~\ref{fig_1}(a). In the ideal case of strong Coulomb
interaction, starting with the polarized state $|+,+\rangle$ one
obtains the maximally entangled state
$\frac{1}{\sqrt{2}}(|+,+\rangle +i |-,-\rangle)$ after an
effective $\pi/2$ two-spin rotation. A longer excitation pulse
results in coherent oscillations between $|+,+\rangle$ and
$|-,-\rangle$ populations --- an experimentally observable
signature of the entanglement between the spins. Schematics of the
pulse sequence and of the evolution of the appropriate dressed
states are shown in Fig.~\ref{fig_1}(b,c). The long optical pulses
used here may be generated by modulating cw lasers, which would
provide sufficiently narrow frequency spectra of the pulses.
Coherent optical coupling of the 5-state system shown in
Fig.~\ref{fig_1}(a) does not yield a dark state, unlike in the
familiar $\Lambda$ system.  However, the two states we use are
long-lived under two-photon resonance, \cite{Hioe} and we can
further reduce trion relaxation by detuning the optical pulses and
by adjusting their amplitudes.

For a single QD in the Voigt configuration with two
single-electron spin states
\begin{equation}
|\pm \rangle =
\frac{1}{\sqrt{2}}(e^\dagger_\downarrow \mp
e^\dagger_\uparrow)|0\rangle,
\end{equation}
we consider only two lowest-energy negative-trion states
\begin{equation}
|t\pm \rangle = \frac{1}{\sqrt{2}}e^\dagger_\downarrow
e^\dagger_\uparrow(h^\dagger_\downarrow \mp
h^\dagger_\uparrow)|0\rangle,
\end{equation}
where the operators $e^\dagger_{\uparrow,\downarrow}$ and
$h^\dagger_{\uparrow,\downarrow}$ create, respectively, an
electron and a heavy hole with spin along or against the growth
direction, which we also take as the optical axis. Because of the
large confinement splitting, the heavy hole is only weakly mixed
with the light hole, and this can be easily compensated for by
adjusting polarizations of the optical fields.\cite{Clive1} With
these restrictions, the system of two dots has 16 states. The four
lowest energy spin states form the qubit sector. They are
separated by a gap from eight single-trion states, which are
similarly distant from four bi-trion states. The interdot Coulomb
interaction of electrons and holes gives rise to a binding energy
of the bi-trion,
\begin{equation}
\Delta=E^{eeee}_{1221}+E^{hhhh}_{1221}-E^{ehhe}_{1221}-E^{ehhe}_{2112},
\end{equation}
where $E^{abba}_{jkkj}$ is a two-particle Coulomb integral, $e$ or
$h$ denotes electron or hole and $j=1,2$ labels the dots, and we
assume that the interdot electron-hole exchange is negligible due
to the large distance. In zero magnetic field, let the transition
energy from the qubit sector to the single-trion sector be
$\omega_{{\rm t}j}$.  The single- to bi-trion transition energy is
shifted by the binding energy $\Delta$, thus enabling the two
types of transition to be independently addressed. Four optical
fields can thus couple the states $|+,- \rangle$ and
$|-,+\rangle$, or states $|+,+\rangle$ and $|-,-\rangle$. In the
following we use the latter pair because an efficient
initialization of the state $|+,+\rangle$ is possible.
\cite{Xiaodong}

Firstly, we develop an analytic model describing the two-qubit
gate. It assumes strong Coulomb interaction between the trions and
does not account for relaxation from the excited states. These
assumptions are relaxed later using numerical simulations of the
system's dynamics.

 The essential process of the quantum operation can be
described by a Hamiltonian
\begin{equation}
 H = \left( {\begin{array}{*{20}c}
   0 &  V_1^*(t) & 0 & 0 & 0\\
   V_1(t) & \delta & H_1^*(t) & 0 & 0 \\
   0 & H_1(t) & 0 & H_2(t) & 0\\
   0 & 0 & H_2^*(t) & \delta & V_2(t) \\
   0 & 0 & 0 & V_2^*(t) & 0
\end{array}} \right), \\
\label{H}
\end{equation}
acting on the five-level system, Fig.~\ref{fig_1}(a), written in
the rotating wave approximation and an interaction picture. The
stationary basis states of the Hamiltonian are $|+,+\rangle$,
$|t+,+\rangle$, $|t+,t-\rangle$, $|-,t-\rangle$, and
$|-,-\rangle$.
The optical fields are detuned by $\delta$ from the single-trion
transitions to avoid populating the intermediate states, while the
two-photon processes are resonant with the bi-trion transition.
For the sake of simplicity we use the same shape for both H-pulses
and both V-pulses. We therefore omit the indices of the pulse envelopes in
Eq.~\ref{H} in the following discussion. The two
H-polarized pulses create the interaction between two dots by
optically coupling the bi-trion state to two single-trion states
in the dots. Then, the shorter V-polarized pulses couple the qubit sector to
the renormalised excited states and rotate the spins in a way
similar to the single qubit operation.\cite{Clive1} The operation
can be described in terms of dressed states, ${\bf C}_{1-5}$. In
the adiabatic approximation for positive $\delta$ their energies
are
\begin{equation}
\begin{array}{*{20}l}
E_1=0, \\
E_{2,3}=\frac{1}{2}(\delta \pm \sqrt{\delta^2+4 V(t)^2}), \\
E_{4,5}=\frac{1}{2}(\delta \pm \sqrt{\delta^2+4 V(t)^2+8 H(t)^2}),
\end{array}
\end{equation}
which are sketched in Fig.~\ref{fig_1}(c). Adiabatic pulses do not
excite transitions to the split-off levels $E_{2,4}$, and thus
states $\textbf{C}_2$, $\textbf{C}_4$ may be ignored. The H-pulse
is applied first and lifts the degeneracy of $E_{1,3}$ and $E_5$
levels, but state ${\bf C}_5$ remains orthogonal to the spin
subspace and thus the initial spin state is not transferred to it.
The transformation of a spin-state is controlled only by the
evolution of the states ${\bf C}_1$ and ${\bf C}_3$, which can be
written as
\begin{equation}
\begin{array}{l}
{\bf C}_1=-\frac{1}{\sqrt{2}}[\cos\theta,0,-\sin\theta,0,\cos\theta], \\
{\bf
C}_3=-\frac{1}{\sqrt{2}}[\cos\varphi_1,-\sin\varphi_1,0,\sin\varphi_1,-\cos\varphi_1],
\end{array}
\label{eig1}
\end{equation}
in terms of time-varying angles defined by
\begin{equation}
\tan \theta = \frac{V(t)}{\sqrt{2}H(t)}, \tan 2\varphi_1
=\frac{2V(t)}{\delta}.
\end{equation}
When the optical fields are switched off, $\textbf{C}_1$ and
$\textbf{C}_3$ reduce to $\frac{1}{\sqrt{2}}[1,0,0,0, \pm 1]$
which belong to the spin sector, ${\bf C}_{2,4}$ to single-trion
states, and ${\bf C}_5$ to $|t+,t-\rangle$.
The evolution of the spin states $|+,+\rangle$ and $|-,-\rangle$
is controlled by the unitary transformation
$e^{-i\phi_1(1-\sigma_x)}$, where
$\sigma_x=|+,+\rangle\langle-,-|+|-,-\rangle\langle+,+|$ and
\begin{equation}
\phi_1=\frac{1}{2}\int E_3(\tau)d\tau,
 \label{phi1}
\end{equation}
where $\hbar=1$ is assumed. An excitation with $\phi_1=\pi/4$
would create a maximally entangled state from either $|+,+\rangle$
or $|-,-\rangle$. The operation is designed to minimize the effects of relaxation
from excited states and pulse imperfections. The states
$\textbf{C}_1$ and $\textbf{C}_3$ overlap within the qubit sector
only. Therefore, the initial state always returns back to the
qubit sector at the end of the operation. If a part of population
is transferred to $\textbf{C}_5$, for example, by applying optical
pulses simultaneously, the bi-trion state will be left populated.
However, this can be minimized by detuning of the two-photon
excitation processes from the bi-trion transitions. Also the
populations of the excited state components of $\textbf{C}_1$ and
$\textbf{C}_3$ are controlled by the small parameters
$(V/\delta)^2$ and $(V/H)^2$. Below we show that it is possible to
maintain the total population of the excited states below $10\%$
for pulse durations of the order of 1~ns. This makes the lifetime
of ${\bf C}_1$ and ${\bf C}_3$ about 10 times longer than that of
bare trions.
For an arbitrary initial state, in addition to two-spin rotation
described above, the $|+,-\rangle$ state acquires a phase
$e^{-i\phi_2}$, where
\begin{equation}
\phi_2=\frac{1}{2}\int[\delta - \sqrt{\delta^2+8 V(\tau)^2}]d\tau,
\label{phi2}
\end{equation}
driven by the V-fields coupling to the single trions
$|t+,-\rangle$ and $|+,t-\rangle$. The optically-induced
transformation of an arbitrary two-spin state in the approximation
of a strong Coulomb coupling and a large splitting between the
Zeeman sublevels is
\begin{equation}
 U_{\rm id} = \left( {\begin{array}{*{20}c}
   e^{-i\phi_1}\cos \phi_1 &  0 & 0 & ie^{-i\phi_1}\sin \phi_1\\
   0 & e^{-i\phi_2} & 0 & 0 \\
   0 & 0 & 1 & 0 \\
   ie^{-i\phi_1}\sin \phi_1 & 0 & 0 & e^{-i\phi_1}\cos \phi_1
\end{array}} \right), \\
\label{U}
\end{equation}
where the phases $\phi_{1,2}$ are defined by Eqs.~(\ref{phi1}) and
(\ref{phi2}) respectively.

\begin{figure}[b]
\centering
\includegraphics[width=0.85\linewidth,  clip=true]{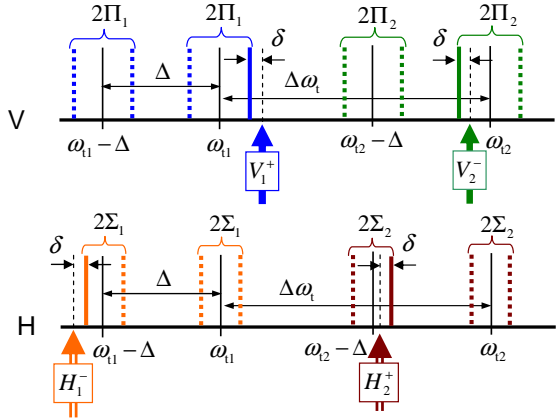}
\caption{(Color online) Energies of allowed optical transitions
versus the optical frequencies (measured in energy units) for
V-polarization (upper figure) and H-polarization (lower figure).
The thin solid lines mark the transition energies in zero magnetic
field. $\omega_{{\rm t}j}$ is the transition energy between a spin
state and  a trion state in  dot $j$. Their difference between the
dots is shown as $\Delta\omega_{\rm t}=\omega_{{\rm
t}2}-\omega_{{\rm t}1}$. $\Delta$ is the bi-trion binding energy,
thus making the transition energy between the single and bi-trion
$\omega_{{\rm t}j}-\Delta$. In a magnetic field, the electron and
hole Zeeman splittings, $\omega^{\rm e}_j$ and $\omega^{\rm h}_j$
in dot $j$, cause the transition energy splitting,
$2\Pi_j=\omega^{\rm e}_j+\omega^{\rm h}_j$ in the V-polarization
and $2\Sigma_j=\omega^{\rm e}_j-\omega^{\rm h}_j$ in the
H-polarization. The Zeeman splitted transitions used in the
quantum operation and off-resonant transitions are denoted by the
thick solid lines and thick dashed lines respectively. The
vertical arrows show the central frequencies of the optical pulses
and their detuning  $\delta$ from the corresponding transitions.}
\label{fig_2}
\end{figure}
Detuning the optical fields is required to avoid unintended
dynamics, such as population transfer from $|+,-\rangle$ to the
single trion states $|t+,-\rangle$ or $|+,t-\rangle$. As an aid to
the design of this process, we gather in Fig.~\ref{fig_2} all the
transition energies for both polarizations. The input parameters
are the energy levels from the dot fabrication, $\Delta$ from dot
placement, the Zeeman splittings, and the central frequencies of
the optical pulses parameterised by single detuning $\delta$ for
simplicity.  Correction operation constrains these parameters as
\begin{equation}
\Delta\omega_{\rm t} \gg \Delta \gg \Pi_i,\Sigma_i \gg \delta,
\label{Cond}
\end{equation}
which is physically reasonable. If the bi-trion binding energy
$\Delta$ and the Zeeman splittings $\Pi_i$ and $\Sigma_i$ are
comparable to the detuning $\delta$, off-resonant processes have
the undesired effect that the pulse sequence which excites the
desired quantum path also excites an path involving the single
trion states $|+,t-\rangle$ and $|t+,-\rangle$, albeit
off-resonantly. This reduces the two-spin rotation angle. This
secondary process can be investigated with a 5-level model similar
to that of the resonant path.  All other off-resonant excitations
just give rise to phases in second-order perturbation.
Including these effects, Eq.~\ref{U}, can
thus be generalized as
\begin{equation}
 U = \left( {\begin{array}{*{20}c}
   e^{-i\phi_{11}}\cos \alpha &  0 & 0 & ie^{-i\phi_{14}}\sin \alpha\\
   0 & e^{-i\phi_{22}} & 0 & 0 \\
   0 & 0 & e^{-i\phi_{33}} & 0 \\
   ie^{-i\phi_{41}}\sin \alpha & 0 & 0 & e^{-i\phi_{44}}\cos \alpha
\end{array}} \right), \\
\label{U1}
\end{equation}
where the phases $\phi_{ij}$ and $\alpha$ are defined in Appendix.
Equation~(\ref{U1}) is not a standard quantum gate. Its usefulness
for quantum information processing has been discussed in
Ref.~\onlinecite{Clive2QD}. In general, the gate can be factorized
as a product of control phase gates and a ${\rm SWAP}$ gate.
Starting with an initially spin-polarized $|+,+\rangle$ or
$|-,-\rangle$ state one can generate a maximally-entangled state
with $\alpha=\pi/4$. Moreover, a longer excitation pulse should
result in coherent two-spin oscillations.

To examine the effects of trion relaxation and off-resonant
pumping, we numerically integrate the equation of motion for the
16-level density matrix including all transitions of
Fig.~\ref{fig_2}. In particular, we consider two
vertically-stacked InAs QDs. We model the trion relaxation with a
Lindblad form,\cite{Clive} and assume that all transitions are
independent with the total relaxation rate $\Gamma=1.2$~$\mu$eV.\cite{Atature}
 The recombination rate of electrons and holes in
different dots, as well as their spin decoherence rate
\cite{Greilich} are negligible on the operation timescale. We take
the interdot difference of the two single-trion energies to be
$\Delta\omega_{\rm t}=10$~meV, and the electron and hole g-factors
to be $g_{\rm e}=-0.48$, $g_{\rm h}=-0.31$ \cite{Xiaodong} for
both dots. There appears to be no experimental data on the
bi-trion binding energy in the literature. Gerardot et al.
obtained 4.56~meV for binding energy of two excitons located in
dots with a vertical separation 4.5~nm.\cite{Gerardot} Scheibner
et al. \cite{Gammon} measured -0.3~meV for the shift of a negative
trion transition when a second dot is occupied by a hole with
respect to a bare transition (interdot distance is 6~nm). These
give us two disparate values for the biexciton binding energy.
From a simple analytical model \cite{Warburton} we estimate
$\Delta=0.8$~meV for dots with vertical separation 8~nm. To
characterize the entanglement of the output qubit state we use the
concurrence, $C$.\cite{Wootters}
\begin{figure}
\centering
\includegraphics[width=0.85\linewidth, clip=true]{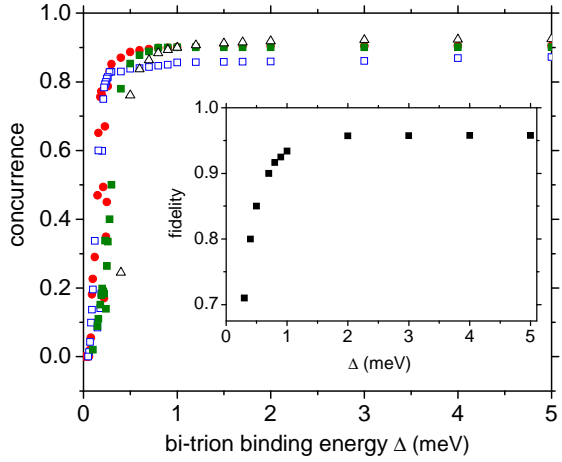}
\caption{(Color online) Concurrence of the output two-spin state
for different bi-trion binding energies. Excitation
parameters: filled circles -- $\delta=-0.1$~meV, $V_0=20$~meV,
$H_0=44$~meV; open squares -- $\delta=-0.13$~meV, $V_0=20$~meV,
$H_0=44$~meV; filled squares -- $\delta=-0.1$~meV, $V_0=10$~meV,
$H_0=44$~meV; open triangles -- $\delta=0.12$~meV, $V_0=15$~meV,
$H_0=65$~meV. $V_0$ and $H_0$ denote amplitudes of the V- and
H-polarized fields. Inset: fidelity of the analytical model
compared to numerical simulations as a function of $\Delta$.}
\label{fig_3}
\end{figure}

The most crucial parameter of the operation is the bi-trion
binding energy $\Delta$. Figure~\ref{fig_3} shows the
concurrence of the output state as a function of
$\Delta$ for several different excitations. The laser
fields are weak enough to avoid unintentional dynamics outside the
16-level system (not studied here). We find that a state with a
concurrence $C>0.85$ can be generated if $\Delta \geq 0.3$~meV for
a broad range of excitation parameters.
%%%SS%%%%%%%%%%%%%%%%%%%%%
The lower boundary for $\Delta$ is determined by the symmetry of
the excitation scheme. One can see in Fig.~\ref{fig_2} that if
$\Delta$ is comparable to the Zeeman splitting the fields $V_2^-$
and $H_2^+$  will excite transitions from the Coulomb-splitt doublets,
in addition to the intended transitions. This effect is
avoided if we design a gate to swap $|+,-\rangle$ and
$|-,+\rangle$ states. In the latter case the concurrence of the
gate remains $C>0.85$ for $\Delta \geq 0.1$~ meV and smoothly
decays to zero at $\Delta \approx 10$~ $\mu$eV.

The time required to entangle two spins is on the order of
fractions of a nanosecond for the whole range of $\Delta$. It is
much shorter than the free-qubit decoherence time ($\sim 1 \mu$sec
) at low temperatures determined by the interaction with a nuclear
spin bath.\cite{Greilich} The main factors limiting the precision
of an operation in this case are excitation of unintended
transitions and relaxation from the optically excited states
utilized in the scheme. Our approach allows precise control for
unintended excitations. Within the 16-level model, if we assume an
infinite relaxation time for the single- and bi-trion states, the
population of the excited states, after the optical fields are
turned off, is less than $10^{-5}$. Variations in pulse shapes or
field intensities do not affect this value. In this sense our
adiabatic excitation scheme is more robust compare to fast
resonant operations utilizing pulse-shaping. Although the effect of
relaxation from the excited states in our scheme is strongly
suppressed by detuning of optical fields it is still noticeable 
and limits the concurrence of a maximally-entangled state. To
further reduce the relaxation effects one has to increase
detunings of optical fields and use QDs with greater separation
between the energy levels (stronger Zeeman splitting and larger
$\Delta$).

To characterize the precision of the designed operation we define a
fidelity of the gate\cite{Poyatos, Clive2QD}
\begin{equation}
F=\overline{\langle \psi_0|(U')^{\dagger} \rho_{\rm f} U'|\psi_0\rangle},
\label{fid}
\end{equation}
as it is described by our adiabatic analytic solution,
Eq.~(\ref{U1}), compare to numerical simulation of quantum
dynamics of the 16-level system that includes non-adiabaticity
effects and relaxation. The bar over Eq.~(\ref{fid}) is for
average over all initial states of two qubits, and $\rho_{\rm f}$
is a two-qubit density matrix obtained in the numerical
simulations. This is the most objective method to analyze the
theoretical model short of having experimental data for
comparison. The inset of Fig.~\ref{fig_3} show that the analytical
model provides a good description of the operation in the same
range of $\Delta$.

\begin{figure}
\centering
\includegraphics[width=0.85\linewidth]{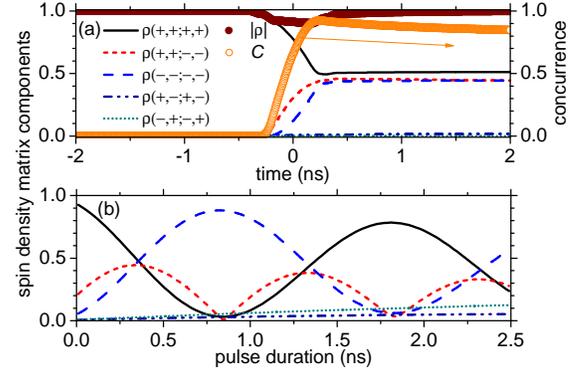}
\caption{(Color online) Evolution of a spin state controlled by
four optical fields. The Coulomb coupling is $\Delta=0.3$~meV, the
detuning is $\delta=-0.1$~meV, the magnetic field is $B=8$~T and
the field amplitudes are $V_0=20$~$\mu$eV and $H_0=44$~$\mu$eV.
Durations of the H- and V-pulses interrelated as $T_H=T_V+2T_{f}$,
$T_{f}=250$~ps is the front duration. (a) Optical pulses are
centered at $t=0$, $T_V=340$~ps. The components not shown in the
figure are below $10^{-3}$ at the end of the excitation. (b) Spin
density matrix as a function of $T_V$.} \label{fig_4}
\end{figure}
\begin{figure}
\centering
\includegraphics[width=0.85\linewidth]{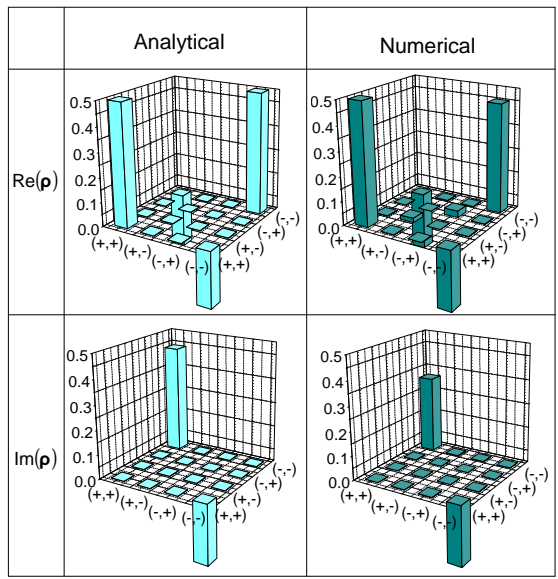}
\caption{(Color online) Density matrix of the output two-qubit state
prepared from $|+,+\rangle$ using an optimised entangling gate.
The analytic solution, obtained using Eq.~\ref{U1}, is compared
with the numerical simulations. Parameters of the dots and the
optical fields are the same as in Fig.~\ref{fig_4}.}\label{fig_5}
\end{figure}

An example of an entangling two-qubit evolution is given in
Fig.~\ref{fig_4} for two dots with the Coulomb coupling
$\Delta=0.3$~meV. The optical pulses, centered at $t=0$, have been
optimized to obtain a final state with a maximal entanglement from
$|+,+\rangle$. The output concurrence $C\approx0.87$ is limited by
relaxation from the single- and bi-trion states. However, because
only a small part of population is transferred to the excited
states the entangling operation is weakly sensitive to the trion
relaxation rate: doubling it results in less than $10\%$ variation
of the concurrence. Longer excitation pulses result in Rabi
oscillations of the pseudo-spin, Fig.~\ref{fig_4}(b), which is
consistent with the analytic model. The decay time of the Rabi
oscillations is of the order of 10~nanosecons. The conventional 3D
tomography plot, Fig.~\ref{fig_5}, shows the two-spin density
matrix after the entangling gate is applied, compared with the ideal
one obtained from Eq.~(\ref{U1}).

To measure the entanglement of the output state in an experiment requires a
full-state tomography,\cite{tomography} which could be rather
difficult and a discussion of which is outside the scope of this work. However partial indication
is provided by the oscillations between states $|+,+\rangle $ and
$|-,- \rangle $ under longer excitation, Fig.~\ref{fig_4}(b). This effect
can be probed by exciting resonantly the population of a given
spin state and then measuring absorption or fluorescence.
With two optical fields, one can selectively excite a transition from a
single two-spin state to a bi-trion state. For instance, optical fields
 $V_1^+$ and $H_2^+$ applied to the systems excite resonantly two-photon
 transition between $|+,+\rangle$ and $|t+,t-\rangle$ states only, see
Fig.~\ref{fig_1}(a). All other transitions are off-resonant. Therefore,
fluorescence should be proportional to the population of $|+,+\rangle$.
To confirm that the fields excite a two-photon transition one could
measure two-photon cross correlations. \cite{Gerardot}

In conclusion, we have developed an adiabatic approach for the
optically-controlled entangling quantum operations with two
electron spins in semiconductor self-assembled quantum dots. The
scheme, utilizing the Coulomb interaction between trions, is
insensitive to material parameters, pulse imperfections and trion
relaxation. We show that using four optical fields a
highly-entangled two-spin state with the concurrence $C>0.85$ can
be prepared on the timescale of the order of 1~ns.

\begin{acknowledgments}
This work was supported by ARO/NSA-LPS and DFG grant BR 1528/5-1.
We thank Dan Gammon, Xiaodong Xu, and Yuli Lyanda-Geller for
helpful discussions.
\end{acknowledgments}

\appendix \section{}

The phases in the transformation matrix, Eq.~\ref{U1}, are defined
as follows:
\begin{align*}
& \alpha = \phi_1+\psi^-, \\
& \phi_{11} = \phi_1+\psi^+ + \int h_1(\tau)d\tau,\\
& \phi_{22} = \phi_2 + \int h_2(\tau)d\tau, \\
& \phi_{33} = \int h_3(\tau)d\tau, \\
& \phi_{44} = \phi_1+\psi^+ + \int h_4(\tau)d\tau, \\
& \phi_{14} = \phi_{41} = \phi_1+\psi^+ + \int h^+(\tau)d\tau,
\end{align*}
where
\begin{align*}
& \psi^{\pm} =(\psi_1 \pm \psi_2)/2,\\
& \psi_1 = -\int\left[\frac{\Delta-\delta}{2}-\sqrt{\frac{(\Delta-\delta)^2}{4}+H^2(\tau)+2V^2(\tau)}\right]d\tau,\\
& \psi_2
=-\int\left[\frac{\Delta-\delta}{2}-\sqrt{\frac{(\Delta-\delta)^2}{4}+H^2(\tau)}\right]d\tau,
\end{align*}
and
\begin{align*}
& h_1(\tau) = -\frac{V^2(\tau)}{2\Pi+\delta}-\frac{H^2(\tau)}{\Delta-\delta+2\Sigma},\\
& h_2(\tau) = \frac{V^2(\tau)}{2\Pi-\delta}-\frac{H^2(\tau)}{\Delta-\delta-2\Sigma}, \\
& h_3(\tau) = -\frac{H^2(\tau)}{\Delta-\delta+2\Sigma}-\frac{H^2(\tau)}{\Delta-\delta-2\Sigma},  \\
& h_4(\tau) =
-\frac{V^2(\tau)}{2\Pi+\delta}+\frac{V^2(\tau)}{2\Pi-\delta},  \\
& h^{\pm}(\tau)=h_1(\tau) \pm h_4(\tau).
\end{align*}

\end{document}